\documentclass[sigconf,  review=false]{acmart}
\usepackage{tikz}
\usepackage{amsmath,amsfonts,amssymb}
\usetikzlibrary{arrows,positioning,shapes.geometric, automata}

\usetikzlibrary{tikzmark}
\usepackage{xpatch}
\expandafter\xpatchcmd
\csname pgfk@/tikz/every picture/.@cmd\endcsname
{\thepage}{\arabic{page}}{}{}

\AtBeginDocument{%
	\providecommand\BibTeX{{%
			\normalfont B\kern-0.5em{\scshape i\kern-0.25em b}\kern-0.8em\TeX}}}

\setcopyright{acmcopyright}
\copyrightyear{2020}
\acmYear{2020}
\acmDOI{10.1145/1122445.1122456}

\acmConference[KDD 2020 Workshop: Designing AI in Support of Good Mental Health (GOOD)]{KDD'20 Designing AI in Support of Good Mental Health (GOOD) Workshop}{August 24th, 2020}{}
\acmBooktitle{KDD'20: KDD GOOD Workshop, August 24th 2020}
\begin{document}
	
	\title{Usable Security for ML Systems in Mental Health:
		A Framework}
	
	
	\author{Helen Jiang}
	%\email{webmaster@marysville-ohio.com}
	\affiliation{%
		\institution{Independent (affiliated with Georgia Institute of Technology)}
	}
	\email{helen.h.jiang@gmail.com}
	
	\author{Erwen Senge}
	\affiliation{
		\institution{Independent}
		%\streetaddress{Tallin, Estonia}
		%\city{Hekla}
		%\country{Iceland}
	}
	\email{erwen@protonmail.com}

	\renewcommand{\shortauthors}{Jiang and Senge}
	
	\begin{abstract}
		While the applications and demands of Machine learning (ML) systems in mental health are growing, there is little discussion nor consensus regarding a uniquely challenging aspect: building security methods and requirements into these ML systems, and keep the ML system usable for end-users. This question of usable security is very important, because the lack of consideration in either security or usability would hinder large-scale user adoption and active usage of ML systems in mental health applications.

		In this short paper, we introduce a framework of four pillars, and a set of desired properties which can be used to systematically guide and evaluate security-related designs, implementations, and deployments of ML systems for mental health. We aim to weave together threads from different domains, incorporate existing views, and propose new principles and requirements, in an effort to lay out a clear framework where criteria and expectations are established, and are used to make security mechanisms usable for end-users of those ML systems in mental health. Together with this framework, we present several concrete scenarios where different usable security cases and profiles in ML-systems in mental health applications are examined and evaluated. 
		
	\end{abstract}
	
	
	\keywords{Mental Health, Machine Learning (ML), Security, Usability, Evaluation, Computer System Life Cycle, Failure Modes}
	
	%% page
	
	\maketitle
	
	\section{Introduction}
	With a mental health crisis looming large and many ML systems being built for mental health use cases, it is challenging to trace, analyze, and compare all the designs and implementations of such systems. So far, there is a lack of well-defined framework that describes properties relating to the security of such ML systems in mental health, and even less considerations are given to how such security mechanisms can be \emph{usable} for those systems' end users. However, without usable security, undiscovered, undisclosed, and ill-considered limitations and properties of security decisions would hold back large-scale adoption and usage\cite{inclu_sec_persuasion} of ML systems in mental health use cases. For more detailed and nuanced discussions, see our treatment at section \ref{diff}. 
	
	The goal of this framework is to establish discussions in communities of mental health, ML, and security, so we can build a common ground for directions and expectations for usable security in ML systems used in mental health scenarios. Moreover, this framework serves to raise awareness, so that both ML and mental health communities will heed this critical aspect of usable security in ML systems for mental health. We hope that this new, interdisciplinary framework would allow researchers and practitioners to systematically compare usable security attributes across ML systems for mental health, meanwhile to identify potential limitations of particular approaches and trade-offs in different scenarios. 

	In this short paper, we propose that ML systems in mental health use cases, beyond the privacy and security requirements already mandated by legislation's and regulations --- for example, Health Insurance Portability and Accountability Act (HIPPA)\cite{hhshippa, hhshippasecrules, nist:criticalinfra} in United States, and General Data Protection Regulation (GDPR) in European Union and its member states' national laws\cite{gdpr:general,gdpr:ehr}  --- should consider properties of usable security proposed by this framework's four pillars, and be evaluated on their (1)\textbf{\textit{context}} models, (2)\textbf{\textit{functionality}} criteria, (3)\textbf{\textit{trustworthiness}} requirements,  and (4)\textbf{\textit{recovery}} principles across their life cycles. 
	
	This work presents our effort to generate discussions and consensus for a common framework in a naturally interdisciplinary area. We built our research on the foundation of computer security research, which has a rich history and long tradition of devising criteria and evaluation rubrics for system designs and implementations. We also incorporated important and recent literature from human-computer interaction (HCI), usable security, and fairness, accountability, and transparency (FAT) research of ML. Weaving these interdisciplinary threads together, we hope that our framework will benefit both researchers and practitioners working on ML systems in mental health. %, especially to bring about best practices.

	\section{Related Work}
	There is a long and distinguished tradition in computer security research: presciently define evaluation criteria and structure assessment frameworks, while research communities were still in their early stages of formation. From this tradition, many remarkable security research outcomes have flourished, and guided the design and building of systems and infrastructure we rely on today\cite{orangebook, iso15408, landwehr1981formal, landwehr2001computer, tpmiso2016, a1answer, ware1970}. However, while the pioneers of security research laid down ``psychological acceptability'' of users as a key principle for secure system design and implementations\cite{saltzer1975protection}, this principle has not been actively researched within the security community until much later while security measures keep confusing even experts\cite{og:zurkoandsimon1996, notenemy:adams1999users, cantencrypt:whitten1999johnny, stillcantencrypt:sheng2006johnny, recipro:chiasson2007even}. Moreover, the ``psychological acceptability'' principle is often doubted as incompatible with the goal of ``security''\cite{og:zurkoandsimon1996, dewitt2006usable, smetters2007usable, theofanos2020usable, soup_sec_comp, usesec:balfanz2004search, usablesec:yee2004aligning, hci:patrick2003hci}, and much of usable security research has traditionally been done in the HCI community, and usable security is still a small community\cite{soups:conf:page, usable_sec_gatech_payne2008brief, usable_sec_nist_all} compared to other areas of security research.
	
	While ``psychological acceptability'' principle is first identified as the meaning of ``usable'' in ``usable security'' \cite{saltzer1975protection, og:zurkoandsimon1996}, there are other efforts trying to precisely define ``usability'' especially in HCI contexts, based on the ``human-centered'' attribute of interactive systems. A prominent example is ISO 9241-210 \cite{iso2019:usability:definition}: ``usability'' is \textsl{``the extent to which a product can be used by specified users to achieve specified goals with effectiveness, efficiency, and satisfaction in a specified context of use''}. Built on this definition, \cite{usable_sec_nist_challenge} of NIST proposed measurements on usability evaluations. 
	
	However, as \cite{og:zurkoandsimon1996, usable_sec_nist_challenge} both point out, measurement of usable security can be highly diverse and context-dependent, meanwhile, such measurements and evaluations focus on the \emph{system} and its interactions with targeted users, often done with small groups in controlled environments\cite{usable_sec_nist_user, usable_1}, with \emph{security} as the users' top concern. While the ``security is top priority'' assumption can be very reasonable for use cases such as national and corporate security, the same assumption likely does not stand when we are evaluating ML systems in mental health use cases, in which users have diverse top priorities. This complicates the already fragmented landscape\cite{recipro:iacono2018consolidating} of usable security, and while ML applications in mental health and FAT ML research are booming\cite{shatte2019machine, xai_framework}, they still do not take usability and security into serious consideration.

	\section{Usable Security Pillars for ML systems}
	Our framework evaluates usable security of ML systems in mental health based on four pillars. Each pillar, in turn, serves as the top concern for each major phase of the computer system life cycle, which can be summarized as: (1)design and implementation; (2)deployment; (3)mass adoption and usage; and finally, (4)maintenance and/or disposal\cite{nist_sys_life_cycle, lifecycle_pendharkar2008empirical, lifecycle_doj}.
	
	\begin{enumerate}
		\item Context: this pillar considers the intended operational environment of the ML system, and how it is designed and built to interact with different types of users with varying purposes, goals, and maliciousness. This pillar is most important during the design and implementation phase of ML systems for mental health.
		
		\item Functionality: this pillar tackles the well-known security-functionality trade-off\cite{og:zurkoandsimon1996, usesec:kainda2010security, to:albrechtsen2010improving, to:hagen2009effects, to:navarro2005approximating}. Keeping ML systems functional while making security usable, it is imperative to ask questions about the complexity and resource-intensity of security methods within the already complex and often resource-intensive ML system, the flexibility of chosen methods to accommodate future security requirements, and how they influence user interactions with the ML system. This pillar is most crucial in the deployment phase of ML systems, especially in the initial stage, when such system is in limited use, without users' significant investment of trust and time. 
		
		\item Trustworthiness: this pillar is by nature user-centered. Many non-expert, lay users are already distrustful and leery of ML, and this set of requirements show that on the matter of security and usability, ML systems may still induce users' trust in the sensitive context of mental health. This pillar is the most critical in achieving active usage and large-scale adoption\cite{inclu_sec_persuasion} of secure ML systems for mental health. %, and how can such trustworthiness be validated
		
		\item Recovery: this pillar handles perhaps one of the toughest challenges in both security and usability: what happens, should a security incident (e.g. a breach, a compromise, or a previously undiscovered vulnerability) happens? What are we going to do with the system and users, now and later? How do we account for the incident this time, to minimize the chance that it would happen again? This pillar is the top priority in maintenance and/or disposal phase of the computer system life cycle.  
	\end{enumerate}

	\subsection{Context Models}
	The list can help ask the right questions for designing and building usable security\cite{a1answer} into ML systems for mental health: it determines what and how much ``usability'' to be considered in a security environment, and move from the more general security threat models, to specific cases of user interactions in mental health scenarios, and also to weigh in negative use cases. The properties below are agnostic to programming languages, software stacks, deployment platforms, and hardware specifications, so they are also flexible enough to accommodate a large class of usable security scenarios for ML systems in mental health. 
	
	\textbf{C1} \emph{Asset Audit.} ML systems in mental health almost inevitably acquire information assets while in use, for example, it may include users' locations, device types etc., as well as patients' functional status information, providers' notes, and organization's intervention plans. Understandably, existing regulations mostly focus on these \emph{acquired} assets. However, in ML systems, ``asset'' is not only acquired, but also \emph{native} to the system itself: its algorithms and models, ground truths, datasets, decision-making logic, and result evaluations, etc. Therefore, identifying both \emph{native} and \emph{acquired} assets of the ML system is critical for usable security. 
	
	\textbf{C2} \emph{Target User Profiling.} ML systems can be utilized by different stakeholders in mental health: from patients, providers, to government officials, they use the system to achieve different goals. Profiling the system's targeted users is the basis to make concrete observations and reasonable estimations, which are then incorporated into design and implementation requirements. Knowing the targeted users and what they use the ML system for, this is usable security's positive case: legitimate users can establish trusted paths and use the system without being hindered by its security requirements. Usable security's negative case is given in \emph{C4}.
	
	\textbf{C3} \emph{Behaviors Categorization.} Behaviors of targeted legitimated users described in \emph{C2} can be either \textit{expected} by the system, or \textit{unexpected} and cause the system to fail, error out, or even trigger security incidents. While it is not possible to iterate through all unexpected behaviors from legitimate users, unexpected user behaviors raise two key components of usable security, and need to be addressed in design and implementation: (1) motivating users to behave in a secure manner so to minimize the systems' failures, errors, and security exposures, because users are not the enemy\cite{notenemy:adams1999users}; (2) when such motivations fail, follow the ``fail-safe'' principle\cite{saltzer1975protection}, meanwhile deliver warning messages about security and failures with usability in mind \cite{warning:sasse2001transforming, warning:sotirakopoulos2011challenges, warning:amran2017usable}. This property is interdependent with \textbf{F5}, where we discuss \emph{robustness}. 
	
	\textbf{C4} \emph{Threat Modeling.} Once the assets are audited and target users and behaviors profiled, threat modeling is essential for security, as threat modeling is a well-studied and used subject in computer security\cite{salter1998toward, shostack2014threat, myagmar2005threat, threatmodel_ms}. There are three main components to consider: (1) \emph{assets} the ML system needs to protect, (2) scope of \emph{interactions} between system and user based on \textbf{C3}; and (3) \emph{malicious actors} and their actions the systems need to defend against. In contrast to \textbf{C2}, \emph{malicious actors} are usable security's negative case: malicious users are stopped or slowed down by the system's security measures.

	\subsection{Functionality Criteria}
	The following properties are most useful when seen from a deployment perspective. They describe \emph{how a ML system works with in-place security requirements while interacting with users.} 
	
	\textbf{F1} \emph{Complexity.} Most, if not all ML systems and applications have at least one of the three constraints: time, memory, and computational power. Therefore, any security measures should consider these constraints and its impact on how well the ML system serves the end users. For example, in a high concurrency event where many users are utilizing the same ML system, if a given security method uses negligible computational power resource on users' end but consumers a lot of system resources, we should consider alternatives for this security method. To measure such complexity, we can use either formal algorithmic complexity notions (e.g. Big O, little O), or empirical evaluations. For example, in 10-user, 100-user, 1,000-user concurrency scenarios, what is the average computational overhead or latency for specific sets of security requirements, with other software and hardware constraints stay the same.

	\textbf{F2} \emph{Availability.} For large-scale ML systems, e.g. mental health use cases with multiple targeted user groups, security measures also need to scale. Availability evaluates how well security methods can generalize to cover a ML system's targeted users and behaviors (\textbf{C2, C3}) without hindering their access to the ML system. A quantitative heuristic for availability of security measures is estimated user adoption rates across user groups, as well as among the genera user base. 

	Notice that the availability criteria is a trade-off to the ``least common mechanism'' principle for secure system design\cite{saltzer1975protection}, and the relative importance between the two are dependent on results of \emph{Context Modeling}, in particular \textbf{C4}. Regardless of which one of the two weighs more heavily in specific scenarios, the security mechanism in question must be carefully designed, judiciously implemented, and rigorously tested before real user runs. 
	
	\textbf{F3} \emph{Flexibility.} Retrofitting security to usability is usually a bad idea and doesn't work well\cite{usesec:balfanz2004search, usablesec:yee2004aligning}, therefore it is important to not only prioritize usable security when designing and building ML systems for mental health, but also to not let current implementations become roadblocks to additional security requirements or system capabilities. Having flexibility accommodates future changes in the system and shifting user base, and is a long-term commitment to the system's usable security traits.
	
	\textbf{F4} \emph{Experience Validation.} To ascertain that security measures did not hold back users, it is crucial to validate real user interactions and experience with the system, regardless of the methods: ideal controlled environments, synthetic experiments, or random sampling. For positive case of usable security (\textbf{C2}) that makes the system more secure but not harder for legitimate users, conducting user studies to evaluate their experiences, interactions, effectiveness, and satisfactions with the system\cite{robust:krol2016towards, hciframework:sweeney1993evaluating} would be indispensable evidence for the ML system's real-world usability. 
	
	\textbf{F5} \emph{Robustness.} Robustness is well-researched in computer system\cite{ieee:robustness, robust:nistiot,robustness:baker2008assessment:robustness, robustness:fernandez2005model:robustness, robustness:sussman2007building}, and recent interests in adversarial ML\cite{robustml:didona2015enhancing, robustml:carlini2017towards, robustml:carlini2019evaluating} has early roots in ML robustness\cite{robustnessml:tcheng1989building:robustness}. In our consideration, robustness is also related to recovery principles in section \ref{recovery_principle}, and has two angles: (1) for \emph{security}, to tolerate and withstand certain errors and faults from the ML layer, the system layer, and user interaction layer; and (2) for \emph{usability}, to communicates to users clearly and timely, when trusted paths cannot be established because of scenarios exceeding (1)'s robustness levels. Interdependent with this criterion is \textbf{C3} for unexpected user behavior categorization. 
	
	\subsection{Trustworthiness Requirements}
	Many non-experts are suspicious and distrustful of ML, because of ML's ``blackbox magic'' reputation. Moreover, the technical nature of FAT ML methods has not endeared lay users towards machine learning either. Now, suppose that another layer of hard-to-use and hard-to-navigate security measures and designs is added to an ML system, such distrust is perhaps only going to grow more intense and open. 
	
	While the users' sentiment of distrust is understandable, the need for good mental health is agnostic about one's feelings towards machine learning and usability of security designs. Therefore, to enable active usage and large-scale adoption\cite{inclu_sec_persuasion} of secure ML systems in mental health cases, it is important to first induce users' trust in the ML systems used, before their active utilization of such ML systems. The trustworthiness requirement suggests \emph{how ML systems in mental health may still earn users' trust, through its security and usability, by well-designed user interactions and communications.} 
	
	\textbf{T1} \emph{Clarity.} Articulating relevant security mechanisms, and their intents, impacts, and implications to users, is fundamental to trust-building. We identified three clarity aspects: (1)clarity of \emph{ML}, where certain artifacts of the ML system's decision-making logic and process (e.g. summary statistics, explanations for classification labels) are exposed and explained to user in \emph{non-technical} manners; (2)clarity of \emph{security}, where user-facing security mechanisms (e.g. trusted path establishment, or revocation of access delegation), and these mechanisms' intents and purposes, are disclosed before users engage in these security mechanisms and take actions, preferably in \emph{non-technical} terms; and (3)clarity of \emph{failure modes}, where recovery (section \ref{recovery_principle}) plan in case of security incidents, is summarized and communicated to users in \emph{non-technical} terminology. 
	
	\textbf{T2} \emph{Constraints.} Complementary to \textbf{T1}, whose focus is on positive cases --- i.e. what can be and is done --- this requirement focuses mostly on negative cases. While providing clarity, ML systems need to draw boundaries and limitations on their capabilities and responsibilities, and then communicate such information. When determining the scope of usable security and communicating to users, we suggest three main factors: (1) limitations, emphasizing what the \emph{system} cannot do (e.g. delegating access without explicit user actions from a trusted path), is not authorized to do (e.g. sharing chatbot history with unknown third parties), or unwilling to do (e.g. exposing ML models' features and parameters) for technical and non-technical reasons; (2) boundaries, concerning what the \emph{user}'s actions cannot accomplish; and (3) expectations, dealing with \emph{interactions} between users and systems, on what users' expectations for the systems should \emph{not} be. This requirement may seem counterintuitive, but it is founded on the ``fail-safe'' principle of computer security\cite{failsafe:cisa,saltzer1975protection}: the default situation is lack of access --- that is, by default, actions and operations are constrained and not allowed to execute. %Research on psychological loss aversion in human behaviors \cite{lossaversion:kahneman1991anomalies,} \cite{lossaversion:quinn2020loss}
	
	\textbf{T3} \emph{Consistency and Stability.} For similar user behaviors under similar contextual conditions, ideally, usability- and security-related experience and interactions should be: (1) similar, within fixed ML systems (data, algorithm, procedure, parameters, input), and (2) comparable, across different ML systems capable to cover the same contextual conditions in their use cases. We name it ``consistency'' property. Conversely, for the same usability and security methods, when provided with the same user behavior inputs, should respond with similar user experience and interaction. We call this ``stability'' property. There properties can help users build their own mental models for how security mechanisms and the general ML system work, and align their expectations with the system's responses. 
	
	Note that we controlled the variables (``similar'', ``fixed'', ``same'') while describing consistency and stability, therefore consistency is not constancy, and stability is not staleness. In fact, the dynamic nature of usable security and the user expectation-system behavior alignment model are both well-known\cite{consistency:garfinkel2003practical}. The goal of alignment, is to motivate secure user behavior and raise user's trust level in the system, and consistency and stability are inroads to alignment.

	\textbf{T4} \emph{Reciprocity.} Leveraging the human tendency to return favors, ML systems in mental health can elicit actions of trust from users, and motivate their secure behaviors and active engagements, as HCI research showed\cite{recipro:fogg1997users, recipro:fogg2002persuasive, recipro:chiasson2007even}: after users receive helpful information from a computer system, they are more likely to provide useful information or actions back to the system. For reciprocity schemes in ML systems in mental health, we identify two stages: (1)\emph{initial exchange} of reciprocity, where after volunteering helpful information to users, the system prompts user for desirable information or behavior input; and (2)\emph{continuous engagement}, meaning that after the initial round, if the user reciprocates, the system should aim to maintain exchanges with users, when user behaviors and other contextual conditions warrant so. Depending on specific areas where the ML system needs to induce trust and motivate behaviors (e.g. having users enable security features, or actively use ML capacities), details of the interaction mechanisms, from the initial offer of help to ongoing engagement patterns, will vary. 
	
	Because reciprocity largely depends on user interactions with the system, it naturally focuses on usability, and has different trade-off with security for different context models. Therefore, any reciprocity schemes must be designed, implemented, and validated judiciously to defend against reciprocity attacks\cite{recipro:attack:zhu2011reciprocity}. 
	
	\subsection{Recovery Principles} \label{recovery_principle}
	Good security needs failure modes, and usable security is no exception. With a variety of assets to protect \textbf{C1}, many functionalities to perform, and user trust to gain and maintain, ML systems in mental health must have a concrete plan for security failures. These principles lay out a foundation to consider \emph{the immediate and long-term aftermath of security incidents and their responses}, so ML systems in mental health can retain usable security attributes and rebuild trust with users (\textbf{T4}). 
	
	\textbf{R1} \emph{Response.} Previous research\cite{data_expo, usable_sec_nist_challenge} surveyed security incidents such as user data leaks, but did not address more complex security challenges to ML systems in mental health, whose sensitive and diverse assets, both native and acquired, make juicy targets. Therefore, ML systems must have protocols and procedures in place, timely reviewed and revised, and ready to respond to security incidents, to achieve three goals: (1) evaluate scope and impact of incident, (2) minimize damages to impacted assets, (3) investigate and attribute sources of incident, and most importantly, (4) rebuild trust in users for the system. (1) through (3) address immediate actions, while (4) is a long-term process that ensures ML systems can maintain its stay with user bases in mental health. This principle is related to \textbf{C1}, \textbf{C4}, and trustworthiness requirements.

	\textbf{R2} \emph{Provenance and Chronology.} The usability of security, in its failure mode, entails that security failures can be traced, examined, analyzed, and inform future security decisions, and such need is satisfied by post-incident provenance and chronology. In ML systems for mental health, provenance and chronology should not only supply (1) a time ordering of system events, technical vulnerabilities or disadvantages, procedural limitations, uncovered edge cases, user interactions, statistics, and likely warning signals leading up to the incident, but also (2) records of any changes (e.g. content, metadata, mode, appearance) in impacted assets (e.g. manipulated ML model parameter, altered user interface, leaked health history), from when the incident \emph{happened}, to when it is \emph{uncovered}. Both provenance and chronology can be considered for user-facing purposes as a tool for repair (\textbf{R3}) and to rebuild trustworthiness.  
	
	\textbf{R3} \emph{Repair.} Post-security-incident repair has two aspects: (1)repairing the system itself, and (2)repairing users' trust in the system. (1) is the direct logical next step of \textbf{R1} and \textbf{R2} with immediate impact and results, while (2) tends to be long-term, and is more difficult --- it needs all the building blocks of trustworthiness to repair users' trust in ML systems impacted by security incidents, especially when incidents concern user data, user-system interaction, or even users' offline behaviors. \emph{Repairing} trust needs to address additional psychological barriers of users, hence harder than \emph{building} trust at first, but it is still possible when \textbf{T2} and \textbf{T4} are emphasized and utilized in the repair process. 
	
	\section{Discussion}
	
	\subsection{Form \& Intent}
	Our framework is a suggestion, an encouragement, a proposal, and an invitation to the community to start acknowledging and researching usability and security in ML systems for mental health. While our framework is not a standardized rubric, we realize that it may become a foundation for future standards, guidelines, or recommendations by organizations such as NIST, ISO, or IEEE, for usable security in generic interactive ML systems, or specifically in mental health applications. Previously, standards were issued on transparency and autonomy in autonomous systems\cite{standard:shahriari2017ieee}, and we are sanguine about a general consensus on usable security in ML systems, especially for mental health use cases. 
	
	\subsection{Scope of Audience \& Usage}
	We intentionally crafted this framework to be agnostic to ML techniques: hence, we can focus on providing a unified structure that is not only comprehensive enough to cover the current interdisciplinary area between traditional computer security, HCI, and ML, but is also flexible enough to accommodate future changes and progresses in these areas. We hope this framework can enable researchers and practitioners to: 
	
	\begin{enumerate}
		\item Identify gaps in security and usability between their theoretical capacities, design variances, actual implementations, and real-world usage patterns; and 
		\item Quickly appraise properties of particular security and usability methods to decide on the most appropriate mechanism for their desired use cases. 
	\end{enumerate}	
	
	In addition, our evaluation framework can be used as a reporting rubric targeting regulators, government officials, and policy makers, so they can quickly get all information in one place, in a clear, structured, and comparable manner. 
	
	\subsection{A Different Kind of Usable Security} \label{diff}
\begin{figure}
		\caption{From security practitioners to end-users: how security mechanisms and experiences are built and delivered in general software systems}		
		\centering
		\begin{tikzpicture} 
		\tikzset{block/.style= {draw, rectangle, align=center,minimum width=1.5cm,minimum height=1cm},
			input/.style={ % requires library shapes.geometric
				draw,
				trapezium,
				minimum width=1.5cm,
				align=left,
				minimum height=1cm
			},
		}
		\node [block] (start) {Security \\ practitioner};
		\node [block, right =1cm of start] (end) {Users \\ (Part of the general population \\ Can usually be described with \\ cohesive \& uniform models \\ for user needs, behaviors, and actions)};
		\path[draw,->] (start) edge (end);
		%(rr) edge (mrr);
		\end{tikzpicture}
		\label{holder:regular}   
	\end{figure}
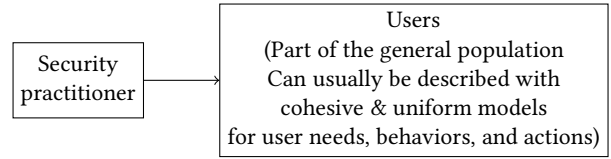

	When we speak of ``practitioners'' in the section above, in the specific context of ML systems for mental health, there are broadly two categories that we target: 
	\begin{enumerate}
		\item Security practitioners: in general system security contexts, security mechanisms and policies are researched, designed, implemented, tested, maintained, and improved by security professionals. % WS2X2x, WS2X
		\item ML practitioners: in general ML system contexts, ML practitioners research, apply, curate, train, validate, test, maintain, improve ML models, algorithms, and date . 
	\end{enumerate}	
	Yet, as we discuss usable security in \emph{ML systems for mental health}, the matter gets more complex: there are more stakeholders, both on the system builders' side, and on the system users' side. And on each side, there are multiple considerations, interests, and mental models that come into play. Table \ref{holder:mlmental} below shows the different stakeholders when we build security to be usable into ML systems for mental health. Comparing it with Figure \ref{holder:regular}, the critical differences between the building usable security into general system versus into ML systems for mental health can be clearly discerned. To summarize: there are more stakeholders on the users' side who deserve usable security for their more diverse needs of the ML system for mental health, and there are more stakeholders on the builders' side who have distinct desires for what they want do with, and how they wish such system to behave. 
	
	For example, while security and ML practitioners desire different ideal attributes from the system and those attributes are not necessarily at odds or contradict with each other, there are trade-offs to make. Between ``strong defense'' with implications for privacy on patient information and ``collect data'' for training when in general, more data is usually better, the builders within themselves need to reach a delicate balance first. On the other end, instead of the cohesive and more-or-less predictable and uniform sets of actions normally expected from user models built for general software or ML systems, we now have a diverse set of potential users, with various sets of actions and behaviors that are not usually taken into account for in those general purpose software or ML systems. In those general purpose systems, Figure \ref{holder:regular} shows the path of how security mechanism and experience are delivered to end users. Behaviors, actions, expectations, and use scenarios of these end users would be captured in user models, and security practitioners would design, build, and deploy security measures and experiences according to those user models. But because of the diverse and varying expectations and actions from distinct groups of end users\footnote{Many ML systems for mental health involve more than one group of stakeholders as shown in Table \ref{holder:mlmental}. In some cases, it might even be possible for one single ML system for mental health to encounter all the four groups of stakeholders. For example, an ML system analyzing facial and verbal expressions in online therapy sessions: the patient and the provider conduct sessions, another caregiver review the analysis after session ends to provide better care, and policy-makers may monitor some sessions for signals of large-scale mental health intervention policies.}, such user models would be too narrow and missing out on legitimate use actions and behaviors. This is a major reason that we crafted this framework: to properly account for and appreciate the diversity and variety of users and their actions in ML systems for mental health, with the end goal to bring a usable and secure experience to all.  
	
	\begin{table*}
		\caption{A sample of stakeholders and their potential needs and purposes when building and using ML systems for mental health. Left of ``||'': builder. Right of ``||'': potential user}
		\label{holder:mlmental}
	
		\begin{tabular}{|p{2.8cm}|p{2.8cm}||p{2.8cm}|p{2.6cm}|p{2.8cm}|p{2.3cm}|}
			\toprule
			\textbf{Security practitioners}&\textbf{ML practitioners}&\textbf{Patients}&\textbf{Providers}&\textbf{Other caregivers}&\textbf{Policy-makers}\\
			\midrule
			Strong defense & Collect data& Get treatment &Diagnose patients&Use by self&Large-scale monitoring\\
			\midrule
			Up-time guarantee & Monitor \& improve models& Get peer support & Treat patients &Use on behalf of patients& Decision-making (e.g. intervention) \\
			\midrule
			Easy rollback & Can validate \& test &Self-monitor& Collaborate with other providers & Assist patients to use& Regulate \\
			\midrule
			East upgrade & Models effective for end-users & Delegate use to other caregivers & Keep records& Monitor patient status& Audit\\
			\midrule
			Good maintenance \& recovery & Deployed model not corrupted & ... &...&...&...\\
			\midrule
			... & ... & ... &...&...&...\\
			\bottomrule
		\end{tabular}
	\end{table*}

	\subsection{Dynamic Relationships}
	As described within sections for each of those sub-attributes, those attributes are not mutually exclusive nor completely independent from each other. Instead, there are rich and dynamic interactions between these sub-attributes, both within a single pillar and across different pillars. Four major types of interactions are list below with short examples.  
	\begin{enumerate}
		\item Inter-dependence: \textbf{F5} and \textbf{C3} are interdependent. In this case, without behavior categorization, robustness is next to impossible to plan for or implement; and without robustness measures tested and used in real-life, it would be very hard to validate if the behavior categorizations are reasonable or sufficient. 
		
		\item Trade-offs: \textbf{F2} is a trade-off to the ``least common mechanism'' principle for secure system design as articulated in \cite{saltzer1975protection}: for security measures to generalize to diverse sets of targeted users and behaviors, commonality increases and distinctions decline. 
		
		\item Prioritization: In many scenarios, prioritizing particular principles before others is the most reasonable and sensible course of action. For example, a large-scale online platform delivering automated conversational therapy may prioritize \textbf{F2} and \textbf{F5}, at the same time de-prioritize trustworthiness requirements based on the assumption that people seeking online automated services generally have greater trust in ML systems and technology, and are likely to be technically proficient enough to navigate security designs built in place. 
		
		\item Complements: \textbf{T1} and \textbf{T2} are complementary: they consider opposite sides of the same issue, and from there, create a comprehensive view and enables balanced and holistic decisions for usable security designs and implementations. 
	\end{enumerate}
	These dynamic and interactive relationships carry deep implications for usable security in ML systems for mental health, and we will explore some examples that showcase these interactive and dynamic relationships between the properties in section \ref{usecase}.
	
	\section{Some examples} \label{usecase}
	We will now apply the four pillared framework, and share several tangible use cases of ML systems for mental health where we evaluate and examine their usable security needs and profiles. This way, we can concretely demonstrate the practicality of our framework, illustrate the dynamic and interactive relationships between the pillars and their corresponding sub-attributes, and showcase the complex and distinct stakeholder demands for usable security that warrant such a framework. We will elaborate on example \ref{case1} with brief comparisons and contrasts to the three other examples, and leave examples \ref{case2} to \ref{case4} for readers' exercise. 
	\begin{enumerate}
		\item Chatbot providing conversation-based therapy to young adults with mental disorders\label{case1}
		\item Auto-diagnosis algorithms of neuro-images for psychiatrists \label{case2}
		\item Personalized matching for providers \& patients \label{case3}
		\item ML system analyzing facial and verbal expressions during tele-therapy sessions \label{case4}
	\end{enumerate}

    In \ref{case1}, the chatbot's context is an online automated services, and hence is more likely to experience high concurrency requests from many different users with existing mental disorders, and such online large-scale services may also be needed in times of distress (e.g. quarantine during COVID-19 global pandemic) to parts of the general population. Therefore, its usable security mechanisms need to prioritize being \emph{available} and \emph{robust} enough to handle more users, and wider ranges behaviors and actions of legitimate users. At the same time, inducing users' \emph{trust} may not be as important, because we may consider people willing to use online chatbot services are more trusting towards ML and technology in general. Although, soon we would see that if security mechanisms and designs are not usable or robust enough, such assumption of trust may not be warranted, and if there were any, would be drastically diminished. 
    
    The assets it needs to protect are not only the security of its general software infrastructure, but also its ML algorithms that generate live conversation responses to users: recall the infamous incident of Microsoft's chatbot Tay on Twitter\cite{mft:tay}, caution must be taken to ensure that legitimate users who need the therapy service could easily and readily access it without much hassle, and that malicious users could be fended off so they could not manipulate the algorithms. While this may seem simple at first, we must properly \emph{categorize behaviors} of our potential legitimate users, who have existing mental disorders. Take attention-deficit/hyperactivity disorder (ADHD) for instance. Suppose that the chatbot implements classes of CAPTCHA or reCAPTCHA methods --- which may include text, image, and sound recognition, as well as text and image matching --- to defend against its \emph{threat model} actors that include bots and malicious users trying to poison its algorithms. While these methods may be effective to defend against these threats, legitimate users with ADHD, whose attention spans are usually shorter than the general population\cite{adhd:nimh}, may be unlikely to complete the CAPTCHAs, especially when there are several ones that come one after another. 
    
    When security defense designs turn legitimate users away, these users may leave with the idea that such mechanisms built to trick them, and the system behind it has no genuine intention to provide them help, and hence their implied trust in the chatbot when they first approached this online automated service, may likely diminish. Hence, it would be advisable to also take \emph{trustworthiness} requirement seriously, especially the \emph{constraint} property. One way to demonstrate it, is using clear language or visual images to inform users of failures. For instance, when a user's attention span is too short to successfully finish a series of reCAPTCHA challenges, a message displays: ``Sorry we could not tell if you are a human or bot. Do you want to try another way?'' This or similar messages could communicate to legitimate users that the system genuinely intends to provide services, the reCAPTCHAs are there because it is a security design, not a farce or trick to turn them away, and there are certain things that these reCAPTCHAs are not capable of doing.  
    
        %%%%%%%%%%%%TODO%%%%%%%%%%%%%%5
    Further thoughts bring more usable security considerations to the discussion on example \ref{case1}. Should the chatbot store any records of its user interactions, so that human providers and caregivers (e.g. parents or legal guardians) of these young adults could monitor their progress, provide better diagnosis, treatment, and care, then we are onto more complex scenarios. There are now two more groups of users to consider, and how to provide usable security for them is a crucial challenge. Moreover, because now there are stored user records, there is an additional \emph{asset} to protect, and the \emph{recovery} principles need to be elevated to higher priorities, especially \emph{repairing} users' trust in the service in the event of a breach or leak. This is where \emph{flexibility} also comes into the picture: if builders of the chatbot had not originally considered this sharing services, and only later decided to add it, \emph{flexibility} of previous security capabilities to accommodate the additional security requirements that come with sharing user records, is extremely important. 
    
    In a base case, even when the chatbot is originally built to not only converse with patients, but also store their records and allows them to share their records with their providers and caregivers, \emph{clarity} of usable security designs that are informed by \emph{behavior categorization} would be integral. For instance, some young adult patients may decide to simply share their passwords with their providers and caretakers for the latter groups to look at their chat records. However, if these patients are in the U.S., this simple act may land their providers and caretakers in legal trouble: because a shared password, even a voluntarily shared one, counts as unauthorized access by the Computer Fraud and Abuse Act\cite{illegal:pswdshare}. Preempting such behaviors would greatly inform usable security decisions when designing and building this chatbot. For example, builders may decide to use methods other than passwords to check for user authorization and authentication status; to utilize the \emph{constraint} criteria, and outwardly warn users to not share their passwords even with trusted providers and caregivers; or to add particular terms in the end-user licensing agreement, security \& privacy policies, or terms \& services documents, and specify cases where the patient could share passwords; or to build a security measure so that patients can delegate access to their records to authenticated providers and caregivers. To choose the most suitable usable security mechanisms or combinations of such mechanisms, builders of the chatbot would need to deliberate on which \emph{contexts}, and especially which \emph{threat models} they decide to focus on. 

    In comparison, example \ref{case2} has a very specific and focused \emph{profile of targeted users} (psychiatrists), so the inter-dependent properties of \emph{behavior categorization} and \emph{robustness} would be straightforward to analyze and design, and the \emph{assets} and \emph{threat models} are also relatively clearly defined and direct. Meanwhile, because such system is used for medical diagnosis, all attributes related to \emph{trustworthiness} need to be \emph{prioritized}: the builders cannot assume that psychiatrists are trusting the ML system's decisions. Moreover, while the \emph{threat models} are relatively simple compared to example \ref{case1} and \ref{case4}, the system still needs to induce trust from the psychiatrists: how do they know it is \emph{them}, instead of \emph{malicious attackers} described in the threat models, who see the images, patients' information, and the algorithm outputs? To address these issues, some security designs may include: ML explanation options that accompany each diagnosis; a side bar that shows access activities; or a device-based two-factor authentication check. Again, it is up to both the ML and security practitioners who build this system, to decide on the specific \emph{contexts} and \emph{threats} they would like to prioritize. 
    	
    Example \ref{case3} also has a straightforward profile for \emph{behaviors} as example \ref{case2}, but the \emph{threat models} could be tricky: because depending on what the builders of the system decide to gather from both the patients and providers for the match, \emph{assets} that the system needs to protect could swing a rather wide range. Meanwhile, because of the usual one-on-one nature of patient-provider relationships, in contrast to example \ref{case1}, de-prioritizing the \emph{availability} of security measures to large numbers of online users could be sensible, and the system might even be able to afford using more time-, memory-, or computationally-\emph{complex} security mechanisms that are nonetheless usable for both providers and patients of the service. For example, incorporating \emph{reciprocity} into the human-system interaction process, by engaging users in short Q\&A games about secure behaviors --- which by the way, could also induces trust from users about the system's security, and fulfill part of the \emph{trustworthiness} requirement. But on the ML front, \emph{trustworthiness} here is similar to the premise of \ref{case1} but the spirit of example \ref{case2}. While patients and providers who choose to use a ML-powered matching service could be assumed to have a greater degree of general trust in ML and technology, the same level of trust could not be assumed in the specific matching decisions: ``How and why did this black-box know that I would be a good fit for this patient/provider?'' would be the question to answer for every patient or provider who uses the service. %On the same front, the system's security measures could also be used users' behaviors  
    
    Example \ref{case4} involves more diverse user groups (patients, providers, likely other caregivers, and potentially policy-makers). Hence, the \emph{assets} need to be protected are more varied and diverse, the \emph{threat models} more complex, the \emph{recovery} scenarios more important, and the usable security mechanisms may need to make trade-offs between \emph{availability} and \emph{flexibility} while still being \emph{functional} when processing live audio and video data, which is another subtle constraint on the \emph{complexity} of usable security mechanisms. Similar to examples \ref{case2} and \ref{case3}, the inevitable question of \emph{trustworthiness} would arise on the ML system's decision rationales and explanations, and there is also an incentive on the builders' end to assure that algorithms and models powering the system are not being manipulated. Because the information being processed and analyzed by the system is largely private and sensitive, convincing different users of the effectiveness and strength of the security mechanisms is also an important task. Comparing it to example \ref{case1} where there are clearer priorities, this system poses a set of full-on challenge for usable security design, implementations, and evaluations. 
    
	\section{Conclusion and Future Work}
	In our work, we presented four categories of desired properties --- based on context, functionality, trustworthiness, and recovery --- to systematically frame and evaluate usable security in ML system for mental health. We discussed those properties' intents, rationales, and sources in the intersection of security, usability, ML, and mental health. We propose that ML systems in mental health be evaluated by the way of this framework for security and usability, in different phases of the computer system life cycle.  
	
	We have analyzed, structured, and presented several examples of ML systems in mental health in this framework, and for next steps, we plan to evaluate more real-life ML systems in mental health, preferably similar to the four described examples, so we can test, validate, and improve our framework and criteria. Simultaneously, we also plan to interview builders of these ML systems in mental health, to understand their awareness of, thought processes behind, and decision rationales of usable security in the systems they designed and built. Because the framework covers the computer system life cycle, while we prefer already deployed, large-scale systems, we are also happy to examine systems in early stages of the cycle. We plan to publish results on websites where this interdisciplinary community can also submit their own framework evaluation results. 
	
	In a deeper dive, our future work will explore a tiered approach to usable security for ML systems, inspired by classic security literature\cite{orangebook}, meanwhile further examine interactions --- e.g. trade-offs, enhancements, overlaps from different perspectives, complements, and interdependence --- between desirable usability and security properties for ML systems in mental health.

	\bibliographystyle{ACM-Reference-Format}
	\bibliography{good_kdd}


\begin{thebibliography}{71}

%%% the Web address.

\ifx \showCODEN    \undefined \def \showCODEN     #1{\unskip}     \fi
\ifx \showDOI      \undefined \def \showDOI       #1{#1}\fi
\ifx \showISBNx    \undefined \def \showISBNx     #1{\unskip}     \fi
\ifx \showISBNxiii \undefined \def \showISBNxiii  #1{\unskip}     \fi
\ifx \showISSN     \undefined \def \showISSN      #1{\unskip}     \fi
\ifx \showLCCN     \undefined \def \showLCCN      #1{\unskip}     \fi
\ifx \shownote     \undefined \def \shownote      #1{#1}          \fi
\ifx \showarticletitle \undefined \def \showarticletitle #1{#1}   \fi
\ifx \showURL      \undefined \def \showURL       {\relax}        \fi
% invisible to TeX
\providecommand\bibfield[2]{#2}
\providecommand\bibinfo[2]{#2}
\providecommand\natexlab[1]{#1}
\providecommand\showeprint[2][]{arXiv:#2}

\bibitem[\protect\citeauthoryear{??}{iee}{1990}]%
        {ieee:robustness}
 \bibinfo{year}{1990}\natexlab{}.
\newblock \showarticletitle{IEEE Standard Glossary of Software Engineering
  Terminology}.
\newblock \bibinfo{journal}{\emph{IEEE Std 610.12-1990}}
  (\bibinfo{year}{1990}), \bibinfo{pages}{1--84}.
\newblock


\bibitem[\protect\citeauthoryear{??}{inc}{2017}]%
        {inclu_sec_persuasion}
 \bibinfo{year}{2017}\natexlab{}.
\newblock \showarticletitle{Inclusive persuasion for security software
  adoption}. In \bibinfo{booktitle}{\emph{Thirteenth Symposium on Usable
  Privacy and Security ({SOUPS} 2017)}}.
\newblock


\bibitem[\protect\citeauthoryear{Adams and Sasse}{Adams and Sasse}{1999}]%
        {notenemy:adams1999users}
\bibfield{author}{\bibinfo{person}{Anne Adams} {and}
  \bibinfo{person}{Martina~Angela Sasse}.} \bibinfo{year}{1999}\natexlab{}.
\newblock \showarticletitle{Users are not the enemy}.
\newblock \bibinfo{journal}{\emph{Commun. ACM}} \bibinfo{volume}{42},
  \bibinfo{number}{12} (\bibinfo{year}{1999}).
\newblock


\bibitem[\protect\citeauthoryear{Albrechtsen and Hovden}{Albrechtsen and
  Hovden}{2010}]%
        {to:albrechtsen2010improving}
\bibfield{author}{\bibinfo{person}{Eirik Albrechtsen} {and}
  \bibinfo{person}{Jan Hovden}.} \bibinfo{year}{2010}\natexlab{}.
\newblock \showarticletitle{Improving information security awareness and
  behaviour through dialogue, participation and collective reflection. An
  intervention study}.
\newblock \bibinfo{journal}{\emph{Computers \& Security}} \bibinfo{volume}{29},
  \bibinfo{number}{4} (\bibinfo{year}{2010}).
\newblock


\bibitem[\protect\citeauthoryear{Amran, Zaaba, Singh, and Marashdih}{Amran
  et~al\mbox{.}}{2017}]%
        {warning:amran2017usable}
\bibfield{author}{\bibinfo{person}{Ammar Amran}, \bibinfo{person}{Zarul~Fitri
  Zaaba}, \bibinfo{person}{Manmeet~Mahinderjit Singh}, {and}
  \bibinfo{person}{Abdalla~Wasef Marashdih}.} \bibinfo{year}{2017}\natexlab{}.
\newblock \showarticletitle{Usable security: Revealing end-users comprehensions
  on security warnings}.
\newblock \bibinfo{journal}{\emph{Procedia Computer Science}}
  \bibinfo{volume}{124} (\bibinfo{year}{2017}).
\newblock


\bibitem[\protect\citeauthoryear{Baker, Schubert, and Faber}{Baker
  et~al\mbox{.}}{2008}]%
        {robustness:baker2008assessment:robustness}
\bibfield{author}{\bibinfo{person}{Jack~W Baker}, \bibinfo{person}{Matthias
  Schubert}, {and} \bibinfo{person}{Michael~H Faber}.}
  \bibinfo{year}{2008}\natexlab{}.
\newblock \showarticletitle{On the assessment of robustness}.
\newblock \bibinfo{journal}{\emph{Structural Safety}} \bibinfo{volume}{30},
  \bibinfo{number}{3} (\bibinfo{year}{2008}).
\newblock


\bibitem[\protect\citeauthoryear{Balfanz, Durfee, Smetters, and
  Grinter}{Balfanz et~al\mbox{.}}{2004}]%
        {usesec:balfanz2004search}
\bibfield{author}{\bibinfo{person}{Dirk Balfanz}, \bibinfo{person}{Glenn
  Durfee}, \bibinfo{person}{Diana~K Smetters}, {and} \bibinfo{person}{Rebecca~E
  Grinter}.} \bibinfo{year}{2004}\natexlab{}.
\newblock \showarticletitle{In search of usable security: Five lessons from the
  field}.
\newblock \bibinfo{journal}{\emph{IEEE Security \& Privacy}}
  \bibinfo{volume}{2}, \bibinfo{number}{5} (\bibinfo{year}{2004}).
\newblock


\bibitem[\protect\citeauthoryear{Carlini, Athalye, Papernot, Brendel, Rauber,
  Tsipras, Goodfellow, Madry, and Kurakin}{Carlini et~al\mbox{.}}{2019}]%
        {robustml:carlini2019evaluating}
\bibfield{author}{\bibinfo{person}{Nicholas Carlini}, \bibinfo{person}{Anish
  Athalye}, \bibinfo{person}{Nicolas Papernot}, \bibinfo{person}{Wieland
  Brendel}, \bibinfo{person}{Jonas Rauber}, \bibinfo{person}{Dimitris Tsipras},
  \bibinfo{person}{Ian Goodfellow}, \bibinfo{person}{Aleksander Madry}, {and}
  \bibinfo{person}{Alexey Kurakin}.} \bibinfo{year}{2019}\natexlab{}.
\newblock \showarticletitle{On evaluating adversarial robustness}.
\newblock \bibinfo{journal}{\emph{arXiv preprint arXiv:1902.06705}}
  (\bibinfo{year}{2019}).
\newblock


\bibitem[\protect\citeauthoryear{Carlini and Wagner}{Carlini and
  Wagner}{2017}]%
        {robustml:carlini2017towards}
\bibfield{author}{\bibinfo{person}{Nicholas Carlini} {and}
  \bibinfo{person}{David Wagner}.} \bibinfo{year}{2017}\natexlab{}.
\newblock \showarticletitle{Towards evaluating the robustness of neural
  networks}. In \bibinfo{booktitle}{\emph{2017 ieee symposium on security and
  privacy (sp)}}. IEEE.
\newblock


\bibitem[\protect\citeauthoryear{Chiasson, van Oorschot, and Biddle}{Chiasson
  et~al\mbox{.}}{2007}]%
        {recipro:chiasson2007even}
\bibfield{author}{\bibinfo{person}{Sonia Chiasson}, \bibinfo{person}{PC van
  Oorschot}, {and} \bibinfo{person}{Robert Biddle}.}
  \bibinfo{year}{2007}\natexlab{}.
\newblock \showarticletitle{Even experts deserve usable security: Design
  guidelines for security management systems}. In
  \bibinfo{booktitle}{\emph{SOUPS Workshop on Usable IT Security Management
  (USM)}}. Citeseer.
\newblock


\bibitem[\protect\citeauthoryear{Commission}{Commission}{2016}]%
        {gdpr:ehr}
\bibfield{author}{\bibinfo{person}{European Commission}.}
  \bibinfo{year}{2016}\natexlab{}.
\newblock \bibinfo{booktitle}{\emph{Overview of the national laws on electronic
  health records in the EU Member States}}.
\newblock
\urldef\tempurl%
\url{https://ec.europa.eu/health/ehealth/projects/nationallaws_electronichealthrecords_en}
\showURL{%
\tempurl}


\bibitem[\protect\citeauthoryear{Commission}{Commission}{2019}]%
        {gdpr:general}
\bibfield{author}{\bibinfo{person}{European Commission}.}
  \bibinfo{year}{2019}\natexlab{}.
\newblock \bibinfo{booktitle}{\emph{Data protection in the EU}}.
\newblock
\urldef\tempurl%
\url{https://ec.europa.eu/info/law/law-topic/data-protection/data-protection-eu_en}
\showURL{%
\tempurl}


\bibitem[\protect\citeauthoryear{Cornejo, Brewer, Edasis, and Piper}{Cornejo
  et~al\mbox{.}}{2016}]%
        {illegal:pswdshare}
\bibfield{author}{\bibinfo{person}{Raymundo Cornejo}, \bibinfo{person}{Robin
  Brewer}, \bibinfo{person}{Caroline Edasis}, {and} \bibinfo{person}{Anne~Marie
  Piper}.} \bibinfo{year}{2016}\natexlab{}.
\newblock \showarticletitle{Vulnerability, sharing, and privacy: Analyzing art
  therapy for older adults with dementia}. In
  \bibinfo{booktitle}{\emph{Proceedings of the 19th ACM Conference on
  Computer-Supported Cooperative Work \& Social Computing}}.
  \bibinfo{pages}{1572--1583}.
\newblock


\bibitem[\protect\citeauthoryear{Cybersecurity and Infrastructure Security
  Agency~(CISA)}{Cybersecurity and Infrastructure Security
  Agency~(CISA)}{2013}]%
        {failsafe:cisa}
\bibfield{author}{\bibinfo{person}{Cybersecurity} {and}
  \bibinfo{person}{United~States Infrastructure Security Agency~(CISA),
  Department of Homeland~Security}.} \bibinfo{year}{2013}\natexlab{}.
\newblock \bibinfo{booktitle}{\emph{Failing Securely}}.
\newblock
\urldef\tempurl%
\url{https://www.us-cert.gov/bsi/articles/knowledge/principles/failing-securely#footnote1_lppmt3d}
\showURL{%
\tempurl}


\bibitem[\protect\citeauthoryear{DeWitt and Kuljis}{DeWitt and Kuljis}{2006}]%
        {dewitt2006usable}
\bibfield{author}{\bibinfo{person}{Alexander~J DeWitt} {and}
  \bibinfo{person}{Jasna Kuljis}.} \bibinfo{year}{2006}\natexlab{}.
\newblock \showarticletitle{Is usable security an oxymoron?}
\newblock \bibinfo{journal}{\emph{interactions}} \bibinfo{volume}{13},
  \bibinfo{number}{3} (\bibinfo{year}{2006}), \bibinfo{pages}{41--44}.
\newblock


\bibitem[\protect\citeauthoryear{Didona, Quaglia, Romano, and Torre}{Didona
  et~al\mbox{.}}{2015}]%
        {robustml:didona2015enhancing}
\bibfield{author}{\bibinfo{person}{Diego Didona}, \bibinfo{person}{Francesco
  Quaglia}, \bibinfo{person}{Paolo Romano}, {and} \bibinfo{person}{Ennio
  Torre}.} \bibinfo{year}{2015}\natexlab{}.
\newblock \showarticletitle{Enhancing performance prediction robustness by
  combining analytical modeling and machine learning}. In
  \bibinfo{booktitle}{\emph{Proceedings of the 6th ACM/SPEC international
  conference on performance engineering}}. \bibinfo{pages}{145--156}.
\newblock


\bibitem[\protect\citeauthoryear{Engineering}{Engineering}{[n.d.]}]%
        {threatmodel_ms}
\bibfield{author}{\bibinfo{person}{Microsoft~Security Engineering}.}
  \bibinfo{year}{[n.d.]}\natexlab{}.
\newblock \bibinfo{booktitle}{\emph{Microsoft Security Development Lifecycle
  (SDL)}}.
\newblock
\urldef\tempurl%
\url{https://www.microsoft.com/en-us/securityengineering/sdl/threatmodeling}
\showURL{%
\tempurl}


\bibitem[\protect\citeauthoryear{Fernandez, Mounier, and Pachon}{Fernandez
  et~al\mbox{.}}{2005}]%
        {robustness:fernandez2005model:robustness}
\bibfield{author}{\bibinfo{person}{Jean-Claude Fernandez},
  \bibinfo{person}{Laurent Mounier}, {and} \bibinfo{person}{Cyril Pachon}.}
  \bibinfo{year}{2005}\natexlab{}.
\newblock \showarticletitle{A model-based approach for robustness testing}. In
  \bibinfo{booktitle}{\emph{IFIP International Conference on Testing of
  Communicating Systems}}. Springer, \bibinfo{pages}{333--348}.
\newblock


\bibitem[\protect\citeauthoryear{Fogg and Nass}{Fogg and Nass}{1997}]%
        {recipro:fogg1997users}
\bibfield{author}{\bibinfo{person}{BJ Fogg} {and} \bibinfo{person}{Clifford
  Nass}.} \bibinfo{year}{1997}\natexlab{}.
\newblock \showarticletitle{How users reciprocate to computers: an experiment
  that demonstrates behavior change}.
\newblock In \bibinfo{booktitle}{\emph{CHI'97 extended abstracts on Human
  factors in computing systems}}.
\newblock


\bibitem[\protect\citeauthoryear{Fogg}{Fogg}{2002}]%
        {recipro:fogg2002persuasive}
\bibfield{author}{\bibinfo{person}{Brian~J Fogg}.}
  \bibinfo{year}{2002}\natexlab{}.
\newblock \showarticletitle{Persuasive technology: using computers to change
  what we think and do}.
\newblock \bibinfo{journal}{\emph{Ubiquity}}  \bibinfo{volume}{2002}
  (\bibinfo{year}{2002}).
\newblock


\bibitem[\protect\citeauthoryear{for Cybersecurity}{for Cybersecurity}{2008}]%
        {iso15408}
\bibfield{author}{\bibinfo{person}{European Union~Agency for Cybersecurity}.}
  \bibinfo{year}{2008}\natexlab{}.
\newblock \bibinfo{booktitle}{\emph{ISO/IEC Standard 15408}}.
\newblock
\urldef\tempurl%
\url{https://www.enisa.europa.eu/topics/threat-risk-management/risk-management/current-risk/laws-regulation/rm-ra-standards/iso-iec-standard-15408}
\showURL{%
\tempurl}


\bibitem[\protect\citeauthoryear{for Standardization}{for
  Standardization}{2016}]%
        {tpmiso2016}
\bibfield{author}{\bibinfo{person}{International~Organization for
  Standardization}.} \bibinfo{year}{2016}\natexlab{}.
\newblock \bibinfo{booktitle}{\emph{ISO/IEC 11889-1:2015 Information technology
  — Trusted platform module library — Part 1: Architecture}}.
\newblock
\urldef\tempurl%
\url{https://www.iso.org/standard/66510.html}
\showURL{%
\tempurl}


\bibitem[\protect\citeauthoryear{for Standardization}{for
  Standardization}{2019}]%
        {iso2019:usability:definition}
\bibfield{author}{\bibinfo{person}{International~Organization for
  Standardization}.} \bibinfo{year}{2019}\natexlab{}.
\newblock \bibinfo{booktitle}{\emph{ISO 9241-210:2019 Ergonomics of
  human-system interaction — Part 210: Human-centred design for interactive
  systems}}.
\newblock
\urldef\tempurl%
\url{https://www.iso.org/standard/77520.html}
\showURL{%
\tempurl}


\bibitem[\protect\citeauthoryear{Garfinkel, Spafford, and Schwartz}{Garfinkel
  et~al\mbox{.}}{2003}]%
        {consistency:garfinkel2003practical}
\bibfield{author}{\bibinfo{person}{Simson Garfinkel}, \bibinfo{person}{Gene
  Spafford}, {and} \bibinfo{person}{Alan Schwartz}.}
  \bibinfo{year}{2003}\natexlab{}.
\newblock \bibinfo{booktitle}{\emph{Practical UNIX and Internet security}}.
\newblock \bibinfo{publisher}{" O'Reilly Media, Inc."}. 6 pages.
\newblock


\bibitem[\protect\citeauthoryear{Hagen and Albrechtsen}{Hagen and
  Albrechtsen}{2009}]%
        {to:hagen2009effects}
\bibfield{author}{\bibinfo{person}{Janne~Merete Hagen} {and}
  \bibinfo{person}{Eirik Albrechtsen}.} \bibinfo{year}{2009}\natexlab{}.
\newblock \showarticletitle{Effects on employees' information security
  abilities by e-learning}.
\newblock \bibinfo{journal}{\emph{Information Management \& Computer Security}}
  (\bibinfo{year}{2009}).
\newblock


\bibitem[\protect\citeauthoryear{Iacono, Smith, von Zezschwitz, Gorski, and
  Nehren}{Iacono et~al\mbox{.}}{2018}]%
        {recipro:iacono2018consolidating}
\bibfield{author}{\bibinfo{person}{Luigi~Lo Iacono}, \bibinfo{person}{Matthew
  Smith}, \bibinfo{person}{Emanuel von Zezschwitz}, \bibinfo{person}{Peter~Leo
  Gorski}, {and} \bibinfo{person}{Peter Nehren}.}
  \bibinfo{year}{2018}\natexlab{}.
\newblock \showarticletitle{Consolidating Principles and Patterns for
  Human-centred Usable Security Research and Development}. In
  \bibinfo{booktitle}{\emph{European Workshop on Usable Security, London}}.
\newblock


\bibitem[\protect\citeauthoryear{Kainda, Flechais, and Roscoe}{Kainda
  et~al\mbox{.}}{2010}]%
        {usesec:kainda2010security}
\bibfield{author}{\bibinfo{person}{Ronald Kainda}, \bibinfo{person}{Ivan
  Flechais}, {and} \bibinfo{person}{AW Roscoe}.}
  \bibinfo{year}{2010}\natexlab{}.
\newblock \showarticletitle{Security and usability: Analysis and evaluation}.
  In \bibinfo{booktitle}{\emph{2010 International Conference on Availability,
  Reliability and Security}}. IEEE.
\newblock


\bibitem[\protect\citeauthoryear{Karunakaran, Thomas, Bursztein, and
  Comanescu}{Karunakaran et~al\mbox{.}}{2018}]%
        {data_expo}
\bibfield{author}{\bibinfo{person}{Sowmya Karunakaran}, \bibinfo{person}{Kurt
  Thomas}, \bibinfo{person}{Elie Bursztein}, {and} \bibinfo{person}{Oxana
  Comanescu}.} \bibinfo{year}{2018}\natexlab{}.
\newblock \showarticletitle{Data Breaches: User Comprehension, Expectations,
  and Concerns with Handling Exposed Data}. In
  \bibinfo{booktitle}{\emph{Fourteenth Symposium on Usable Privacy and Security
  ({SOUPS} 2018)}}.
\newblock


\bibitem[\protect\citeauthoryear{Krol, Spring, Parkin, and Sasse}{Krol
  et~al\mbox{.}}{2016}]%
        {robust:krol2016towards}
\bibfield{author}{\bibinfo{person}{Kat Krol}, \bibinfo{person}{Jonathan~M
  Spring}, \bibinfo{person}{Simon Parkin}, {and} \bibinfo{person}{M~Angela
  Sasse}.} \bibinfo{year}{2016}\natexlab{}.
\newblock \showarticletitle{Towards robust experimental design for user studies
  in security and privacy}. In \bibinfo{booktitle}{\emph{The $\{$LASER$\}$
  Workshop: Learning from Authoritative Security Experiment Results
  ($\{$LASER$\}$ 2016)}}.
\newblock


\bibitem[\protect\citeauthoryear{Landwehr}{Landwehr}{1981}]%
        {landwehr1981formal}
\bibfield{author}{\bibinfo{person}{Carl~E Landwehr}.}
  \bibinfo{year}{1981}\natexlab{}.
\newblock \showarticletitle{Formal models for computer security}.
\newblock \bibinfo{journal}{\emph{ACM Computing Surveys (CSUR)}}
  \bibinfo{volume}{13}, \bibinfo{number}{3} (\bibinfo{year}{1981}),
  \bibinfo{pages}{247--278}.
\newblock


\bibitem[\protect\citeauthoryear{Landwehr}{Landwehr}{2001}]%
        {landwehr2001computer}
\bibfield{author}{\bibinfo{person}{Carl~E Landwehr}.}
  \bibinfo{year}{2001}\natexlab{}.
\newblock \showarticletitle{Computer security}.
\newblock \bibinfo{journal}{\emph{International Journal of Information
  Security}} \bibinfo{volume}{1}, \bibinfo{number}{1} (\bibinfo{year}{2001}),
  \bibinfo{pages}{3--13}.
\newblock


\bibitem[\protect\citeauthoryear{Mohamed and Chiasson}{Mohamed and
  Chiasson}{2018}]%
        {usable_1}
\bibfield{author}{\bibinfo{person}{Reham~Ebada Mohamed} {and}
  \bibinfo{person}{Sonia Chiasson}.} \bibinfo{year}{2018}\natexlab{}.
\newblock \showarticletitle{Online Privacy and Aging of Digital Artifacts}. In
  \bibinfo{booktitle}{\emph{Fourteenth Symposium on Usable Privacy and Security
  ({SOUPS} 2018)}}.
\newblock


\bibitem[\protect\citeauthoryear{Myagmar, Lee, and Yurcik}{Myagmar
  et~al\mbox{.}}{2005}]%
        {myagmar2005threat}
\bibfield{author}{\bibinfo{person}{Suvda Myagmar}, \bibinfo{person}{Adam~J
  Lee}, {and} \bibinfo{person}{William Yurcik}.}
  \bibinfo{year}{2005}\natexlab{}.
\newblock \showarticletitle{Threat modeling as a basis for security
  requirements}. In \bibinfo{booktitle}{\emph{Symposium on requirements
  engineering for information security (SREIS)}}, Vol.~\bibinfo{volume}{2005}.
  Citeseer.
\newblock


\bibitem[\protect\citeauthoryear{National Institute~of Mental~Healtn}{National
  Institute~of Mental~Healtn}{2019}]%
        {adhd:nimh}
\bibfield{author}{\bibinfo{person}{NIH National Institute~of Mental~Healtn}.}
  \bibinfo{year}{2019}\natexlab{}.
\newblock \bibinfo{booktitle}{\emph{Attention-Deficit/Hyperactivity Disorder}}.
\newblock
\urldef\tempurl%
\url{https://www.nimh.nih.gov/health/topics/attention-deficit-hyperactivity-disorder-adhd/index.shtml}
\showURL{%
\tempurl}


\bibitem[\protect\citeauthoryear{Navarro and Foley}{Navarro and Foley}{2005}]%
        {to:navarro2005approximating}
\bibfield{author}{\bibinfo{person}{Guillermo Navarro} {and}
  \bibinfo{person}{Simon~N Foley}.} \bibinfo{year}{2005}\natexlab{}.
\newblock \showarticletitle{Approximating SAML using similarity based
  imprecision}. In \bibinfo{booktitle}{\emph{International Conference on
  Intelligence in Communication Systems}}. Springer, \bibinfo{pages}{191--200}.
\newblock


\bibitem[\protect\citeauthoryear{NIST}{NIST}{[n.d.]}]%
        {robust:nistiot}
\bibfield{author}{\bibinfo{person}{NIST}.} \bibinfo{year}{[n.d.]}\natexlab{}.
\newblock \bibinfo{booktitle}{\emph{Network security \& robustness}}.
\newblock
\urldef\tempurl%
\url{https://www.nist.gov/topics/network-security-robustness}
\showURL{%
\tempurl}


\bibitem[\protect\citeauthoryear{of~Defense}{of~Defense}{1985}]%
        {orangebook}
\bibfield{author}{\bibinfo{person}{Department of Defense}.}
  \bibinfo{year}{1985}\natexlab{}.
\newblock \bibinfo{booktitle}{\emph{Trusted Computer System Evaluation
  Criteria}}.
\newblock
\urldef\tempurl%
\url{https://csrc.nist.gov/csrc/media/publications/conference-paper/1998/10/08/proceedings-of-the-21st-nissc-1998/documents/early-cs-papers/dod85.pdf}
\showURL{%
\tempurl}


\bibitem[\protect\citeauthoryear{of~Health and Services}{of~Health and
  Services}{2014}]%
        {hhshippasecrules}
\bibfield{author}{\bibinfo{person}{United States~Department of Health} {and}
  \bibinfo{person}{Human Services}.} \bibinfo{year}{2014}\natexlab{}.
\newblock \bibinfo{booktitle}{\emph{{HIPAA Security Rule Crosswalk to NIST
  Cybersecurity Framework}}}.
\newblock
\urldef\tempurl%
\url{https://www.hhs.gov/sites/default/files/nist-csf-to-hipaa-security-rule-crosswalk-02-22-2016-final.pdf}
\showURL{%
\tempurl}


\bibitem[\protect\citeauthoryear{of~Justice Systems Development Life Cycle
  Guidance~Document}{of~Justice Systems Development Life Cycle
  Guidance~Document}{2003}]%
        {lifecycle_doj}
\bibfield{author}{\bibinfo{person}{The~Department of Justice Systems
  Development Life Cycle Guidance~Document}.} \bibinfo{year}{2003}\natexlab{}.
\newblock \bibinfo{booktitle}{\emph{THE SYSTEM DEVELOPMENT LIFE CYCLE (SDLC)}}.
\newblock
\urldef\tempurl%
\url{https://www.justice.gov/archive/jmd/irm/lifecycle/table.htm}
\showURL{%
\tempurl}


\bibitem[\protect\citeauthoryear{of~Standards and Technology}{of~Standards and
  Technology}{[n.d.]}]%
        {nist_sys_life_cycle}
\bibfield{author}{\bibinfo{person}{National~Institute of Standards} {and}
  \bibinfo{person}{Technology}.} \bibinfo{year}{[n.d.]}\natexlab{}.
\newblock \bibinfo{booktitle}{\emph{THE SYSTEM DEVELOPMENT LIFE CYCLE (SDLC)}}.
\newblock
\urldef\tempurl%
\url{https://csrc.nist.gov/csrc/media/publications/shared/documents/itl-bulletin/itlbul2009-04.pdf}
\showURL{%
\tempurl}


\bibitem[\protect\citeauthoryear{of~Standards and Technology}{of~Standards and
  Technology}{2017a}]%
        {usable_sec_nist_all}
\bibfield{author}{\bibinfo{person}{National~Institute of Standards} {and}
  \bibinfo{person}{Technology}.} \bibinfo{year}{2017}\natexlab{a}.
\newblock \bibinfo{booktitle}{\emph{Usable Cybersecurity}}.
\newblock
\urldef\tempurl%
\url{https://csrc.nist.gov/Projects/Usable-Cybersecurity}
\showURL{%
\tempurl}


\bibitem[\protect\citeauthoryear{of~Standards and Technology}{of~Standards and
  Technology}{2017b}]%
        {usable_sec_nist_user}
\bibfield{author}{\bibinfo{person}{National~Institute of Standards} {and}
  \bibinfo{person}{Technology}.} \bibinfo{year}{2017}\natexlab{b}.
\newblock \bibinfo{booktitle}{\emph{Usable Cybersecurity: Behavior}}.
\newblock
\urldef\tempurl%
\url{https://csrc.nist.gov/Topics/Security-and-Privacy/security-and-behavior/behavior}
\showURL{%
\tempurl}


\bibitem[\protect\citeauthoryear{of~Standards and Technology}{of~Standards and
  Technology}{2014}]%
        {nist:criticalinfra}
\bibfield{author}{\bibinfo{person}{National~Institute of Standards} {and}
  \bibinfo{person}{United~States Technology}.} \bibinfo{year}{2014}\natexlab{}.
\newblock \bibinfo{booktitle}{\emph{Framework for Improving Critical
  Infrastructure Cybersecurity}}.
\newblock
\urldef\tempurl%
\url{https://www.nist.gov/system/files/documents/cyberframework/cybersecurity-framework-021214.pdf}
\showURL{%
\tempurl}


\bibitem[\protect\citeauthoryear{Patrick, Long, and Flinn}{Patrick
  et~al\mbox{.}}{2003}]%
        {hci:patrick2003hci}
\bibfield{author}{\bibinfo{person}{Andrew~S Patrick}, \bibinfo{person}{A~Chris
  Long}, {and} \bibinfo{person}{Scott Flinn}.} \bibinfo{year}{2003}\natexlab{}.
\newblock \showarticletitle{HCI and security systems}. In
  \bibinfo{booktitle}{\emph{CHI'03 Extended Abstracts on Human Factors in
  Computing Systems}}.
\newblock


\bibitem[\protect\citeauthoryear{Payne and Edwards}{Payne and Edwards}{2008}]%
        {usable_sec_gatech_payne2008brief}
\bibfield{author}{\bibinfo{person}{Bryan~D Payne} {and}
  \bibinfo{person}{W~Keith Edwards}.} \bibinfo{year}{2008}\natexlab{}.
\newblock \showarticletitle{A brief introduction to usable security}.
\newblock \bibinfo{journal}{\emph{IEEE Internet Computing}}
  \bibinfo{volume}{12}, \bibinfo{number}{3} (\bibinfo{year}{2008}),
  \bibinfo{pages}{13--21}.
\newblock


\bibitem[\protect\citeauthoryear{Pendharkar, Rodger, and
  Subramanian}{Pendharkar et~al\mbox{.}}{2008}]%
        {lifecycle_pendharkar2008empirical}
\bibfield{author}{\bibinfo{person}{Parag~C Pendharkar},
  \bibinfo{person}{James~A Rodger}, {and} \bibinfo{person}{Girish~H
  Subramanian}.} \bibinfo{year}{2008}\natexlab{}.
\newblock \showarticletitle{An empirical study of the Cobb--Douglas production
  function properties of software development effort}.
\newblock \bibinfo{journal}{\emph{Information and Software Technology}}
  \bibinfo{volume}{50}, \bibinfo{number}{12} (\bibinfo{year}{2008}),
  \bibinfo{pages}{1181--1188}.
\newblock


\bibitem[\protect\citeauthoryear{Qin, Lapets, Jansen, Flockhart, Albab,
  Globus-Harris, Roberts, and Varia}{Qin et~al\mbox{.}}{2019}]%
        {soup_sec_comp}
\bibfield{author}{\bibinfo{person}{Lucy Qin}, \bibinfo{person}{Andrei Lapets},
  \bibinfo{person}{Frederick Jansen}, \bibinfo{person}{Peter Flockhart},
  \bibinfo{person}{Kinan~Dak Albab}, \bibinfo{person}{Ira Globus-Harris},
  \bibinfo{person}{Shannon Roberts}, {and} \bibinfo{person}{Mayank Varia}.}
  \bibinfo{year}{2019}\natexlab{}.
\newblock \showarticletitle{From Usability to Secure Computing and Back Again}.
  In \bibinfo{booktitle}{\emph{Fifteenth Symposium on Usable Privacy and
  Security ({SOUPS} 2019)}}.
\newblock


\bibitem[\protect\citeauthoryear{Salter, Saydjari, Schneier, and
  Wallner}{Salter et~al\mbox{.}}{1998}]%
        {salter1998toward}
\bibfield{author}{\bibinfo{person}{Chris Salter}, \bibinfo{person}{O~Sami
  Saydjari}, \bibinfo{person}{Bruce Schneier}, {and} \bibinfo{person}{Jim
  Wallner}.} \bibinfo{year}{1998}\natexlab{}.
\newblock \showarticletitle{Toward a secure system engineering methodolgy}. In
  \bibinfo{booktitle}{\emph{Proceedings of the 1998 workshop on New security
  paradigms}}.
\newblock


\bibitem[\protect\citeauthoryear{Saltzer and Schroeder}{Saltzer and
  Schroeder}{1975}]%
        {saltzer1975protection}
\bibfield{author}{\bibinfo{person}{Jerome~H Saltzer} {and}
  \bibinfo{person}{Michael~D Schroeder}.} \bibinfo{year}{1975}\natexlab{}.
\newblock \showarticletitle{The protection of information in computer systems}.
\newblock \bibinfo{journal}{\emph{Proc. IEEE}} (\bibinfo{year}{1975}).
\newblock


\bibitem[\protect\citeauthoryear{Sasse, Brostoff, and Weirich}{Sasse
  et~al\mbox{.}}{2001}]%
        {warning:sasse2001transforming}
\bibfield{author}{\bibinfo{person}{Martina~Angela Sasse},
  \bibinfo{person}{Sacha Brostoff}, {and} \bibinfo{person}{Dirk Weirich}.}
  \bibinfo{year}{2001}\natexlab{}.
\newblock \showarticletitle{Transforming the ‘weakest link’—a
  human/computer interaction approach to usable and effective security}.
\newblock \bibinfo{journal}{\emph{BT technology journal}} \bibinfo{volume}{19},
  \bibinfo{number}{3} (\bibinfo{year}{2001}).
\newblock


\bibitem[\protect\citeauthoryear{{Schaefer}}{{Schaefer}}{2004}]%
        {a1answer}
\bibfield{author}{\bibinfo{person}{M. {Schaefer}}.}
  \bibinfo{year}{2004}\natexlab{}.
\newblock \showarticletitle{If A1 is the answer, what was the question? An Edgy
  Naif's retrospective on promulgating the trusted computer systems evaluation
  criteria}. In \bibinfo{booktitle}{\emph{20th Annual Computer Security
  Applications Conference}}. \bibinfo{pages}{204--228}.
\newblock


\bibitem[\protect\citeauthoryear{Shahriari and Shahriari}{Shahriari and
  Shahriari}{2017}]%
        {standard:shahriari2017ieee}
\bibfield{author}{\bibinfo{person}{Kyarash Shahriari} {and}
  \bibinfo{person}{Mana Shahriari}.} \bibinfo{year}{2017}\natexlab{}.
\newblock \showarticletitle{IEEE standard review—Ethically aligned design: A
  vision for prioritizing human wellbeing with artificial intelligence and
  autonomous systems}. In \bibinfo{booktitle}{\emph{2017 IEEE Canada
  International Humanitarian Technology Conference (IHTC)}}. IEEE,
  \bibinfo{pages}{197--201}.
\newblock


\bibitem[\protect\citeauthoryear{Shatte, Hutchinson, and Teague}{Shatte
  et~al\mbox{.}}{2019}]%
        {shatte2019machine}
\bibfield{author}{\bibinfo{person}{Adrian~BR Shatte}, \bibinfo{person}{Delyse~M
  Hutchinson}, {and} \bibinfo{person}{Samantha~J Teague}.}
  \bibinfo{year}{2019}\natexlab{}.
\newblock \showarticletitle{Machine learning in mental health: a scoping review
  of methods and applications}.
\newblock \bibinfo{journal}{\emph{Psychological medicine}}
  \bibinfo{volume}{49}, \bibinfo{number}{9} (\bibinfo{year}{2019}),
  \bibinfo{pages}{1426--1448}.
\newblock


\bibitem[\protect\citeauthoryear{Sheng, Broderick, Koranda, and Hyland}{Sheng
  et~al\mbox{.}}{2006}]%
        {stillcantencrypt:sheng2006johnny}
\bibfield{author}{\bibinfo{person}{Steve Sheng}, \bibinfo{person}{Levi
  Broderick}, \bibinfo{person}{Colleen~Alison Koranda}, {and}
  \bibinfo{person}{Jeremy~J Hyland}.} \bibinfo{year}{2006}\natexlab{}.
\newblock \showarticletitle{Why johnny still can’t encrypt: evaluating the
  usability of email encryption software}. In
  \bibinfo{booktitle}{\emph{Symposium On Usable Privacy and Security}}. ACM.
\newblock


\bibitem[\protect\citeauthoryear{Shostack}{Shostack}{2014}]%
        {shostack2014threat}
\bibfield{author}{\bibinfo{person}{Adam Shostack}.}
  \bibinfo{year}{2014}\natexlab{}.
\newblock \bibinfo{booktitle}{\emph{Threat modeling: Designing for security}}.
\newblock \bibinfo{publisher}{John Wiley \& Sons}.
\newblock


\bibitem[\protect\citeauthoryear{Smetters}{Smetters}{2007}]%
        {smetters2007usable}
\bibfield{author}{\bibinfo{person}{D Smetters}.}
  \bibinfo{year}{2007}\natexlab{}.
\newblock \bibinfo{title}{Usable security: Oxymoron or challenge}.
\newblock
\newblock


\bibitem[\protect\citeauthoryear{Sokol and Flach}{Sokol and Flach}{2020}]%
        {xai_framework}
\bibfield{author}{\bibinfo{person}{Kacper Sokol} {and} \bibinfo{person}{Peter
  Flach}.} \bibinfo{year}{2020}\natexlab{}.
\newblock \showarticletitle{Explainability Fact Sheets: A Framework for
  Systematic Assessment of Explainable Approaches}. In
  \bibinfo{booktitle}{\emph{Proceedings of the 2020 Conference on Fairness,
  Accountability, and Transparency}} \emph{(\bibinfo{series}{FAT* ’20})}.
\newblock
\urldef\tempurl%
\url{https://doi.org/10.1145/3351095.3372870}
\showURL{%
\tempurl}


\bibitem[\protect\citeauthoryear{Sotirakopoulos, Hawkey, and
  Beznosov}{Sotirakopoulos et~al\mbox{.}}{2011}]%
        {warning:sotirakopoulos2011challenges}
\bibfield{author}{\bibinfo{person}{Andreas Sotirakopoulos},
  \bibinfo{person}{Kirstie Hawkey}, {and} \bibinfo{person}{Konstantin
  Beznosov}.} \bibinfo{year}{2011}\natexlab{}.
\newblock \showarticletitle{On the challenges in usable security lab studies:
  lessons learned from replicating a study on SSL warnings}. In
  \bibinfo{booktitle}{\emph{Proceedings of the Seventh Symposium on Usable
  Privacy and Security}}.
\newblock


\bibitem[\protect\citeauthoryear{Sussman}{Sussman}{2007}]%
        {robustness:sussman2007building}
\bibfield{author}{\bibinfo{person}{Gerald~Jay Sussman}.}
  \bibinfo{year}{2007}\natexlab{}.
\newblock \showarticletitle{Building robust systems an essay}.
\newblock \bibinfo{journal}{\emph{Citeseer}}  \bibinfo{volume}{113}
  (\bibinfo{year}{2007}), \bibinfo{pages}{1324}.
\newblock


\bibitem[\protect\citeauthoryear{Sweeney, Maguire, and Shackel}{Sweeney
  et~al\mbox{.}}{1993}]%
        {hciframework:sweeney1993evaluating}
\bibfield{author}{\bibinfo{person}{Marian Sweeney}, \bibinfo{person}{Martin
  Maguire}, {and} \bibinfo{person}{Brian Shackel}.}
  \bibinfo{year}{1993}\natexlab{}.
\newblock \showarticletitle{Evaluating user-computer interaction: a framework}.
\newblock  (\bibinfo{year}{1993}).
\newblock


\bibitem[\protect\citeauthoryear{Tcheng, Lambert, Lu, and Rendell}{Tcheng
  et~al\mbox{.}}{1989}]%
        {robustnessml:tcheng1989building:robustness}
\bibfield{author}{\bibinfo{person}{David Tcheng}, \bibinfo{person}{Bruce
  Lambert}, \bibinfo{person}{Stephen~CY Lu}, {and} \bibinfo{person}{Larry
  Rendell}.} \bibinfo{year}{1989}\natexlab{}.
\newblock \showarticletitle{Building robust learning systems by combining
  induction and optimization}.
\newblock \bibinfo{journal}{\emph{Urbana}}  \bibinfo{volume}{100}
  (\bibinfo{year}{1989}), \bibinfo{pages}{61801}.
\newblock


\bibitem[\protect\citeauthoryear{Theofanos}{Theofanos}{2015}]%
        {usable_sec_nist_challenge}
\bibfield{author}{\bibinfo{person}{Mary Theofanos}.}
  \bibinfo{year}{2015}\natexlab{}.
\newblock \bibinfo{booktitle}{\emph{Usable security}}.
\newblock
\urldef\tempurl%
\url{https://csrc.nist.gov/CSRC/media/Presentations/Usable-Security/images-media/day2_research_430-530pt1.pdf}
\showURL{%
\tempurl}


\bibitem[\protect\citeauthoryear{Theofanos}{Theofanos}{2020}]%
        {theofanos2020usable}
\bibfield{author}{\bibinfo{person}{Mary Theofanos}.}
  \bibinfo{year}{2020}\natexlab{}.
\newblock \showarticletitle{Is Usable Security an Oxymoron?}
\newblock \bibinfo{journal}{\emph{Computer}} \bibinfo{volume}{53},
  \bibinfo{number}{2} (\bibinfo{year}{2020}).
\newblock


\bibitem[\protect\citeauthoryear{{United States Department of Health and Human
  Services}}{{United States Department of Health and Human Services}}{2017}]%
        {hhshippa}
\bibfield{author}{\bibinfo{person}{{United States Department of Health and
  Human Services}}.} \bibinfo{year}{2017}\natexlab{}.
\newblock \bibinfo{title}{Health Information Privacy, {HIPPA for
  Professionals}}.
\newblock
\newblock
\newblock
\shownote{\url{https://www.hhs.gov/hipaa/for-professionals/security/guidance/cybersecurity/index.html}.}


\bibitem[\protect\citeauthoryear{USENIX}{USENIX}{2015}]%
        {soups:conf:page}
\bibfield{author}{\bibinfo{person}{USENIX}.} \bibinfo{year}{2015}\natexlab{}.
\newblock \bibinfo{booktitle}{\emph{USENIX SOUPS 2020 conference page}}.
\newblock
\urldef\tempurl%
\url{https://www.usenix.org/conference/soups2019}
\showURL{%
\tempurl}


\bibitem[\protect\citeauthoryear{Victor}{Victor}{2016}]%
        {mft:tay}
\bibfield{author}{\bibinfo{person}{Daniel Victor}.}
  \bibinfo{year}{2016}\natexlab{}.
\newblock \bibinfo{booktitle}{\emph{Microsoft Created a Twitter Bot to Learn
  From Users. It Quickly Became a Racist Jerk.}}
\newblock
\urldef\tempurl%
\url{https://www.nytimes.com/2016/03/25/technology/microsoft-created-a-twitter-bot-to-learn-from-users-it-quickly-became-a-racist-jerk.html}
\showURL{%
\tempurl}


\bibitem[\protect\citeauthoryear{Ware}{Ware}{1970}]%
        {ware1970}
\bibfield{author}{\bibinfo{person}{Willis Ware}.}
  \bibinfo{year}{1970}\natexlab{}.
\newblock \bibinfo{booktitle}{\emph{Security Controls for Computer Systems:
  Report of Defense Science Board Task Force on Computer Security}}.
\newblock
\urldef\tempurl%
\url{https://csrc.nist.gov/csrc/media/publications/conference-paper/1998/10/08/proceedings-of-the-21st-nissc-1998/documents/early-cs-papers/ware70.pdf}
\showURL{%
\tempurl}


\bibitem[\protect\citeauthoryear{Whitten and Tygar}{Whitten and
  Tygar}{[n.d.]}]%
        {cantencrypt:whitten1999johnny}
\bibfield{author}{\bibinfo{person}{Alma Whitten} {and} \bibinfo{person}{J~Doug
  Tygar}.} \bibinfo{year}{[n.d.]}\natexlab{}.
\newblock \showarticletitle{Why Johnny Can't Encrypt: A Usability Evaluation of
  PGP 5.0.}
\newblock


\bibitem[\protect\citeauthoryear{Yee}{Yee}{2004}]%
        {usablesec:yee2004aligning}
\bibfield{author}{\bibinfo{person}{Ka-Ping Yee}.}
  \bibinfo{year}{2004}\natexlab{}.
\newblock \showarticletitle{Aligning security and usability}.
\newblock \bibinfo{journal}{\emph{IEEE Security \& Privacy}}
  \bibinfo{volume}{2}, \bibinfo{number}{5} (\bibinfo{year}{2004}).
\newblock


\bibitem[\protect\citeauthoryear{Zhu, Carpenter, Kulkarni, and Kolimi}{Zhu
  et~al\mbox{.}}{2011}]%
        {recipro:attack:zhu2011reciprocity}
\bibfield{author}{\bibinfo{person}{Feng Zhu}, \bibinfo{person}{Sandra
  Carpenter}, \bibinfo{person}{Ajinkya Kulkarni}, {and} \bibinfo{person}{Swapna
  Kolimi}.} \bibinfo{year}{2011}\natexlab{}.
\newblock \showarticletitle{Reciprocity attacks}. In
  \bibinfo{booktitle}{\emph{Proceedings of the Seventh Symposium on Usable
  Privacy and Security}}.
\newblock


\bibitem[\protect\citeauthoryear{Zurko and Simon}{Zurko and Simon}{1996}]%
        {og:zurkoandsimon1996}
\bibfield{author}{\bibinfo{person}{Mary~Ellen Zurko} {and}
  \bibinfo{person}{Richard~T. Simon}.} \bibinfo{year}{1996}\natexlab{}.
\newblock \showarticletitle{User-Centered Security}. In
  \bibinfo{booktitle}{\emph{Proceedings of the 1996 Workshop on New Security
  Paradigms}} \emph{(\bibinfo{series}{NSPW ’96})}.
  \bibinfo{publisher}{Association for Computing Machinery}.
\newblock


\end{thebibliography}

\end{document}